\documentstyle[eqsecnum,aps]{revtex}
\begin{document}
%\tightenlines
%\draft

\newcommand{\be}{\begin{eqnarray}}
\newcommand{\ee}{\end{eqnarray}}
\newcommand{\dia}{\!\!\!\!\!\!\!\not\,\,\,\,}
\newcommand{\nn}{\nonumber}

\title{Longitudinal and transverse  fermion-boson vertex in QED at
       finite temperature in the HTL approximation}
\author{Alejandro Ayala}
\address{Instituto de Ciencias Nucleares       
         Universidad Nacional Aut\'onoma de M\'exico\\
         Apartado Postal 70-543, M\'exico Distrito Federal 04510, M\'exico.}
\author{Adnan Bashir}
\address{Instituto de F{\'\i}sica y Matem\'aticas
         Universidad Michoacana de San Nicol\'as de Hidalgo\\
         Apartado Postal 2-82, Morelia, Michoac\'an 58040, M\'exico.}
\maketitle
\begin{abstract}
We evaluate the fermion-photon vertex in QED at the one loop level in Hard 
Thermal Loop approximation and  write it in covariant form. The complete
vertex can be expanded in terms of 32 basis vectors. As is well known,
the fermion-photon vertex and the fermion propagator are related
through a Ward-Takahashi Identity (WTI). This relation splits the
vertex into two parts: longitudinal ($\Gamma_L$) and transverse
($\Gamma_T$). $\Gamma_L$ is fixed by the WTI. The description  of
the longitudinal part consumes 8 of the basis vectors. The remaining
piece $\Gamma_T$ is then written in terms of 24 spin
amplitudes. Extending the work of Ball and Chiu and
K{\i}z{\i}lers\"{u} {\it et. al.}, we propose a set of basis 
vectors $T^{\mu}_i(P_1,P_2)$ at finite temperature such that each of these is 
transverse to the photon four-momentum and also satisfies
$T^{\mu}_i(P,P)=0$, in accordance with the Ward Identity, with their
corresponding coefficients being free of kinematic singularities. This basis
reduces to the form proposed by K{\i}z{\i}lers\"{u} {\it et. al.} at zero
temperature. We also evaluate explicitly the coefficient of each of
these vectors at the above-mentioned level of approximation.
\end{abstract}

\section{Introduction}

Schwinger-Dyson equations provide a natural starting point to study 
the non-perturbative aspects  of gauge field  theories, in particular in
connection  with dynamical  symmetry breaking. As is well known, knowledge 
of the three-point vertex is crucial in such studies.
Numerous works exist at zero temperature with a gradual improvement 
on the choice of the full vertex and other underlying 
assumptions, rendering the technique increasingly powerful. 
At finite temperature, the complexity of the equations and the number of the 
unknown functions to be evaluated grow enormously. For example, in QED at
zero temperature, the three-point vertex can be written in terms of 12 spin
amplitudes, whereas, at finite temperature, this number becomes
32, almost a three-fold increase in the number of unknowns involved.
As a result, most of the research is restricted to the use of the 
bare vertex~\cite{Dorey1,Dorey2,Aitchison2,Aitchison3,Triantaphyllou1,Triantaphyllou2,Lee1}. 

However, the Ward-Takahashi Identity (WTI) inspired vertex has
more recently been employed to study the dynamical breakdown of chiral
symmetry \cite{strickland}. The WTI relates the three-point vertex to
the fermion propagator. This identity permits us to decompose the full vertex
into two parts, the longitudinal part, which is fixed by the WTI~\cite{BC},
and the transverse part, which vanishes on contraction with the
boson momentum and hence remains undetermined by the WTI. For instance 
in reference~\cite{strickland}, Strickland uses the longitudinal part 
of the vertex to study, among other
things, the dynamically generated mass as a function of temperature. 
He finds out that the gauge dependence of the critical temperature
for which chiral symmetry is restored reduces considerably as compared
to the results obtained earlier using the bare vertex. 

The lessons learned from similar studies at zero temperature suggest that 
the complete gauge independence of the critical temperature relies crucially 
on the choice of the transverse vertex. In this context, perturbation theory 
is an important point of reference as it is natural to believe that physically
meaningful solutions of the Schwinger-Dyson equations must agree with
perturbative results in the weak coupling regime. This realization 
has been exploited in Refs.~\cite{CP,BP1,BKP1,BR1} to derive constraints
on the fermion propagator and the three-point vertex.

In this paper, we extend the works in Refs.~\cite{BC,KRP,devy1,devy2}
from zero to finite temperature, evaluating the one-loop transverse
vertex in the Hard Thermal Loop (HTL) approximation and writing it in
covariant form. The procedure we use is as follows: By evaluating the
fermion  propagator to a given order, it is possible to determine the
longitudinal vertex to the same order. A subtraction of the
longitudinal part then yields the transverse part, the one which is
not fixed by the WTI. According to the choice of Ball and Chiu, which
was later modified by K{\i}z{\i}lers\"{u} {\it et. al.}~\cite{KRP},
the transverse vertex can be expanded out, at zero temperature, in terms
of 8 independent spin structures. We find out that at finite
temperature the transverse vertex requires 24 spin structures to be expressed.
Following a systematic transition from zero to finite temperature,
we construct a basis of 24 transverse vectors. We use this basis to 
write out the fermion-photon vertex in the HTL
approximation~\cite{Braaten}, which is known to be a gauge independent
scheme valid for temperatures much larger than the fermion mass. The
results allow us to make useful comments on the possible
non-perturbative structures of the vertex at finite temperature. The
work is organized as follows: In Sect.~\ref{II}, we briefly recall the
expressions for the fermion self-energy and fermion-photon vertex at
finite temperature in the HTL approximation. In Sect.~\ref{III} we
construct the longitudinal and transverse vertices defining a set of
24 spin structures to serve as the transverse basis. In Sect.~\ref{IV}
we find the explicit expressions for the coefficients
of the transverse vertex. We finally summarize and conclude in
Sect.~\ref{V}. A short appendix collects some of the integrals that
are involved in the computations in Sect.~\ref{IV}.   

\section{Fermion propagator and Fermion-Photon vertex in the HTL
         approximation}\label{II}

We start by expanding out the fermion propagator $S(P)$ in its most general
form at finite temperature. This can be obtained by noticing that the
only available Lorentz structures are $1, P\dia, U\dia, P\dia  U\dia$.
Therefore, we can write (hereafter, we use capital letters 
to refer to four-vectors, whereas lower case letters are used to refer
to their components)  
\be
   S(P)^{-1}&=& (1-a) \, P\dia \, - \, b \, U\dia \, + \, c \, P\dia \, U\dia 
                 \, + \, d     \nonumber\\
       &=& P\dia \, - \, \Sigma(P) \, + \, c \, P\dia \, U\dia \, + \, d  \, ,
   \label{ferprop}
\ee
where 
\be
 \Sigma(P) =  a \, P\dia \, + \, b \, U\dia \;.  \label{ferpropHTL}
\ee
The Lorentz invariant functions $a$,$b$,$c$ and $d$ will in general
depend on two Lorentz scalars $P^2$ and $U \cdot P$,
where $U^\mu$ is the four-velocity of the heat bath as seen from a
general frame. We can choose these scalars to be  
\be
   p_0&\equiv & P^\mu U_\mu\nonumber\\
   p&\equiv &[(P^\mu U_\mu )^2-P^2]^{1/2} \,.
   \label{escalars}
\ee
When the temperature $T$ is high enough, all the $T=0$ masses  can safely be 
neglected in comparison with the scale $gT$ --where
$g$ is the fermion-gauge boson coupling constant. In this regime,
the dominant contributions to n-point functions at one loop come from
the situation in which the loop momentum is {\it hard}, {\it i.e.} of
the order of $T$~\cite{Braaten}. This is the so called HTL approximation 
which is known to render the results gauge independent when applied, for
example, to the computation of damping rates and energy losses of
particles propagating through hot plasmas~\cite{Braaten2}. For QED,
in this approximation, the temperature dependent
$\Sigma (P)$ is given in Minkowski space explicitly by
\be
   \Sigma (P)= m_f^2 \int\frac{d\Omega}{4\pi}
   \frac{\hat{K}\dia}{P\cdot\hat{K}}\, ,
   \label{SigmaInt}
\ee
where $m_f^2=g^2T^2/8$ is the square of the fermion thermal mass and
$\hat{K}^\mu=(-1,{\mathbf\hat{k}})$.
Evaluation of Eq.~(\ref{SigmaInt}) yields the
functions $a$ and $b$ defined in Eq.~(\ref{ferpropHTL})~\cite{Weldon}
\be
   a(p_0,p)&=&-\frac{m_f^2}{p^2}\left[1-p_0L(P)\right]
   \equiv m_f^2 \tilde{a}(p_0,p)\nonumber\\
   b(p_0,p)&=&\frac{m_f^2}{p^2}\left[p_0+\left(p^2-p_0^2
   \right)L(P)\right]\equiv m_f^2 \tilde{b}(p_0,p)\, ,
   \label{abHTL} 
\ee
where
\be
   L(P)\equiv - \int\frac{d\Omega}{4\pi}\frac{1}{P\cdot\hat{K}}
   =\frac{1}{2p}\ln\left[\frac{p_0+p}{p_0-p}\right]\, .
   \label{funcL}
\ee
Equations~(\ref{ferprop}), ~(\ref{ferpropHTL}), ~(\ref{abHTL})
and~(\ref{funcL}) together with $c=d=0$ constitute the QED fermion
propagator at one loop level in the HTL approximation. $d=0$ corresponds
to the case of exact chiral symmetry whereas $c=0$ is required for a
parity conserving theory~\cite{Weldon}. 

As for the temperature dependent fermion-photon vertex
$\Gamma_\mu(P_1,P_2)$, its corresponding HTL expression in Minkowski
space is given by~\cite{LeBellac}  
\be
   \Gamma_\mu(P_1,P_2)=m_f^2\, G_{\mu\nu}(P_1,P_2)\; \gamma^\nu\, ,
   \label{Vert}
\ee
where 
\be
   G_{\mu\nu}(P_1,P_2)\equiv\int\frac{d\Omega}{4\pi}
   \frac{\hat{K}_\mu\hat{K}_\nu}{(P_1\cdot\hat{K})
   (P_2\cdot\hat{K})}        \;.
   \label{funcG}
\ee

\section{Longitudinal and transverse vertex}\label{III}

\subsection{Longitudinal vertex}

The WTI relates the fermion propagator and the fermion-photon vertex
by 
\be
     Q^\mu\Gamma_\mu(P_1,P_2)&=&S^{-1}(P_1)-S^{-1}(P_2) \nn \\
   &=&  \left[1-a(P_1)\right] \, P\dia_1 \, - \, \left[1-a(P_2)\right]
        \, P\dia_2  \, - \, b(P_1) \, U\dia \, + \, b(P_2) \, U\dia \, \nn \\
   &+&  \, c(P_1)  \, P\dia_1 \, U\dia  \,-  \, c(P_2) 
        \, P\dia_2 \, U\dia  \, + \, d(P_1) \, - \, d(P_2)\, ,  \label{WTI}  
\ee
with $Q=P_1-P_2$.
This relation allows us to decompose the
vertex into longitudinal $\Gamma_{\mu}^{L}(P_1,P_2)$ and transverse 
$\Gamma_{\mu}^{T}(P_1,P_2)$ parts 
\be
   \Gamma_{\mu}(P_1,P_2)=\Gamma_{\mu}^{L}(P_1,P_2)+
   \Gamma_{\mu}^{T}(P_1,P_2)\, ,
   \label{decompose} 
\ee
where the transverse part satisfies
\be
   Q^{\mu}\Gamma_{\mu}^{T}(P_1,P_2)=0\;\;\;\;\;\mbox{and} \;\;\;\;
   \Gamma_{\mu}^{T}(P,P)=0\, ,
   \label{transv}
\ee
and  hence remains  undetermined  by WTI. In order to construct the 
longitudinal vertex along the lines of the calculation of Ball and
Chiu~\cite{BC}, we start from the Ward-Identity
\be
  \Gamma_{\mu}^{L}(P,P) = \frac{\partial}{\partial P^{\mu}} \,
  S^{-1}(P)\, ,
 \label{WI}
\ee        
which is the limiting form of WTI when $P_1=P_2=P$. Making use of 
Eq.~(\ref{ferprop}), the above relation yields 
\be
 \Gamma_{\mu}^L(P,P) &=& \left[1-a(P)\right] \, \gamma_{\mu} \, - \,
2 P_{\mu}  P\dia \, \frac{\partial}{\partial P^2} \, a(P) \, - \, 
2 P_{\mu} U\dia  \,  \frac{\partial}{\partial P^2} \, b(P) \nn \\
&+& \, \gamma_{\mu} U\dia c(P) \, + \, 2 P_{\mu} P\dia U\dia \,
 \frac{\partial}{\partial P^2} \, c(P) \, + \, 2 P_{\mu}   \, 
\frac{\partial}{\partial P^2} \, d(P) \;.
\ee
We now symmetrize this expression under the exchange of $P_1 \leftrightarrow
P_2$ in the following manner
\be
   P_{\mu} &\rightarrow& \frac{1}{2} \, \left( P_{1\mu} + P_{2\mu} \right)
\nn  \\
 P\dia &\rightarrow& \frac{1}{2} \, \left( {P\dia}_1 + {P\dia}_2 \right)
\nn  \\
 \frac{\partial}{\partial P^2} \, f(P)  &\rightarrow& 
\frac{f(P_1)-f(P_2)}{P_1^2-P_2^2} \quad,
\ee
where $f(P)$ is a generic function standing for $a(P)$, $b(P)$, $c(P)$ and
$d(P)$. We thus arrive at the non-perturbative expression for the longitudinal
vertex
\be
   \Gamma_{\mu}^{L}(P_1,P_2)&=& \gamma_{\mu} \, - \, \frac{1}{2}
   \left[a(P_1)+a(P_2)\right]\gamma_\mu \, -  \, 
   \frac{a(P_1)-a(P_2)}{2(P_1^2-P_2^2)}
     \left(P_{1\mu}+P_{2\mu}\right)
   \left(P_1\!\dia\,+P_2\!\dia\,\,\right)
  \, -  \, \frac{b(P_1)-b(P_2)}{(P_1^2-P_2^2)}   
   \left(P_{1\mu}+P_{2\mu}\right)U\dia   \nn   \\
&+&  \, \frac{1}{2}   \left[c(P_1)+c(P_2)\right]\gamma_\mu U\dia \, + \,
     \frac{c(P_1)-c(P_2)}{2(P_1^2-P_2^2)}
     \left(P_{1\mu}+P_{2\mu}\right)
   \left(P_1\!\dia\,+P_2\!\dia\,\,\right) U\dia \, + \,
 \, \frac{d(P_1)-d(P_2)}{(P_1^2-P_2^2)}   
   \left(P_{1\mu}+P_{2\mu}\right) \;.
   \label{long}
\ee
Note that the coefficients of two spin structures corresponding to 
$P_{1\mu} P_{2\nu} \sigma^{\mu\nu}$ and $ U\dia P_{1\mu} P_{2\nu} 
\sigma^{\mu\nu}$ will be zero because these do not
appear on the right hand side of Eq.~(\ref{WTI}).
By construction, the longitudinal part of the vertex is free of kinematical 
singularities. An interesting and useful comparison with the work of
Ball an Chiu~\cite{BC} reveals that for every structure $V_{\mu}$
at zero temperature, there exists an additional structure
$V_{\mu} U\dia$ at finite temperature. Therefore,
\be
{\rm zero \quad temperature} \qquad  &\rightarrow& \qquad  {\rm finite \quad 
temperature}   \nn  \\
\gamma_{\mu} \qquad   &\rightarrow&  \qquad \gamma_{\mu} U\dia \nn \\
 \left(P_{1\mu}+P_{2\mu}\right)
   \left(P_1\!\dia\,+P_2\!\dia\,\,\right)  \qquad   &\rightarrow&  \qquad 
 \left(P_{1\mu}+P_{2\mu}\right)
   \left(P_1\!\dia\,+P_2\!\dia\,\,\right) U\dia    \nn  \\
 \left(P_{1\mu}+P_{2\mu}\right)  \qquad   &\rightarrow&  \qquad
  \left(P_{1\mu}+P_{2\mu}\right)U\dia \, .
\ee
This observation would be of help to construct the transverse vertex later
in this section. In the HTL approximation, the one loop longitudinal vertex 
reduces to
\be
   \Gamma_{\mu}^{L}(P_1,P_2)&=&   - \, \frac{1}{2}
   \left[a(P_1) + a(P_2) \right]
\gamma_\mu \, -  \, \left[
   \frac{a(P_1)-a(P_2)}{2(P_1^2-P_2^2)} \right]
     \left(P_{1\mu}+P_{2\mu}\right)
   \left(P_1\!\dia\,+P_2\!\dia\,\,\right)
  \,   \nn \\
&-& \left[ \frac{b(P_1)-b(P_2)}{(P_1^2-P_2^2)} \right]   
   \left(P_{1\mu}+P_{2\mu}\right)U\dia    \quad.
   \label{longHTL}
\ee

\subsection{Transverse Vertex}

\noindent
At finite temperature, the available vectors to expand the full vertex
$\Gamma_{\mu}(P_1,P_2)$ are $\gamma_{\mu},P_{1\mu},P_{2\mu}$ and $U_{\mu}$.
Correspondingly there are 8 Lorentz scalars $1,{\not\! P_1},{\not\! P_2},
{\not\! U},{\not\! P_1}{\not\! P_2},{\not\! P_1}{\not\! U},
{\not\! P_2}{\not\! U},{\not\! P_1}{\not\! P_2}{\not\! U}$. Therefore,
there are 32 spin amplitudes in total: 
\begin{eqnarray}
V_{1\mu}&=&{P_{1\mu}}{\not \! P_1}\qquad , \qquad
V_{2\mu}={P_{2\mu}}{\not \! P_2}  \qquad , \qquad
V_{3\mu}={P_{1\mu}}{\not \! P_2}  \qquad , \qquad
V_{4\mu}={P_{2\mu}}{\not \! P_1}\nonumber\\
V_{5\mu}&=&{\gamma_{\mu}}{\not \! P_1}{\not \! P_2} \quad \, \,, \qquad
V_{6\mu}={\gamma_{\mu}}           \qquad \quad \; \;\,,\qquad
V_{7\mu}={P_{1\mu}}\qquad \quad \;,\qquad
V_{8\mu}={P_{2\mu}}\nonumber\\
V_{9\mu}&=&{P_{1\mu}}{\not \! P_1}{\not \! P_2} \;\;\;,\quad \; \;
V_{10\mu}={P_{2\mu}}{\not \! P_1}{\not \! P_2} \;\;\;, \quad \; \;
V_{11\mu}={\gamma_{\mu}}{\not \! P_1}\qquad \, \, \, \,,\quad \; \, \,
V_{12\mu}={\gamma_{\mu}}{\not \! P_2}\nonumber\\
V_{13\mu}&=&{V_{1\mu}}{\not \! U} \qquad \, \, \, ,\quad \, \, \, 
V_{14\mu}={V_{2\mu}}{\not \! U} \qquad \, \, , \quad \, \, \, \,
V_{15\mu}={V_{3\mu}}{\not \! U}\qquad \,\,\,,\quad \; \;
V_{16\mu}={V_{4\mu}}{\not \! U}\nonumber\\
V_{17\mu}&=&{V_{5\mu}}{\not \! U} \qquad \, \, \,,\quad \, \, \, 
V_{18\mu}={V_{6\mu}}{\not \! U} \qquad \, \, , \quad \, \, \,
V_{19\mu}={V_{7\mu}}{\not \! U}\qquad \;\;\,,\quad \; \;
V_{20\mu}={V_{8\mu}}{\not \! U}\nonumber\\
V_{21\mu}&=&{V_{9\mu}}{\not \! U} \qquad \, \, \, ,\quad \, \, \, 
V_{22\mu}={V_{10\mu}}{\not \! U} \qquad  , \quad \, \, \,
V_{23\mu}={V_{11\mu}}{\not \! U}\qquad \;,\quad \; \;
V_{24\mu}={V_{12\mu}}{\not \! U}\nonumber\\
V_{25\mu}&=&{U_{\mu}} \qquad \quad \, \, \,\,,\quad \, \, \, 
V_{26\mu}={U_{\mu}}{\not \! U} \qquad  \,\,\,\,\,, \quad \, \, \,
V_{27\mu}={U_{\mu}}{\not \! P_1}\qquad \, \, \, \,,\quad \; \;
V_{28\mu}={U_{\mu}}{\not \! P_2}\nonumber\\
V_{29\mu}&=&{U_{\mu}}{\not \! P_1}{\not \! U} \quad \,\,\,\,,\quad \, \, \, 
V_{30\mu}={U_{\mu}}{\not \! P_2}{\not \! U} \quad \,\, \, , \quad \, \, \, \, 
\,
V_{31\mu}={U_{\mu}}{\not \! P_1}{\not \! P_2} \quad \,,\quad \; \;
V_{32\mu}={U_{\mu}}{\not \! P_1}{\not \! P_2}{\not \! U}\, .  \label{vbasis}
\end{eqnarray}
In a construction parallel to that of Ball and Chiu, Eq.~(\ref{long})
consumes 8 of these 32 structures. The transverse vertex
${\Gamma_{\mu}^{T}(P_1,P_2)}$ can then be written in terms of the
remaining 24 basis vectors as 
\be
   \Gamma_{\mu}^{T}(P_1,P_2)=m_f^2\sum_{i=1}^{24} 
   \tau^{i}(P_1^2,P_2^2,Q^2)T_{\mu}^{i}(P_1,P_2)\, ,
   \label{trsnbas}
\ee
where the coefficients $\tau^{i}$ are to be found from the explicit
expression for $\Gamma_{\mu}$ and the choice of the transverse
basis of tensors $\left\{T_{\mu}^{i}\right\}$, which will be a linear
combination of the vectors in Eq.({\ref{vbasis}}). The procedure for their
selection is as follows:
\begin{itemize}

\item  Motivated from the success of the basis proposed by 
K{\i}z{\i}lers\"{u} {\it et. al.}, which is a simple modification of the
one proposed by Ball and Chiu, we choose 8 of the basis vectors the same.

\item Let us now recall the observation made in connection with the 
construction of the longitudinal vertex. We noticed that for every 
structure $V_{\mu}$ at zero temperature, there exists an additional 
structure $V_{\mu} {\not \! U}$ at finite temperature. Making use of this
observation, we select 8 of the basis vectors, which are a simple 
multiplication of the previous 8 vectors by ${\not \! U}$. Note that a mere
multiplication with ${\not \! U}$ does not change the much desired 
transversality 
properties of the vectors and the fact that they vanish for $P_1=P_2$.

\item
The remaining 8 transverse vectors must contain the vector $U_{\mu}$ not
used so far. Now the transversality of the new vectors can be guaranteed
if they have the form  
\be
\nn (Q\cdot U)R_i V_{i\mu} + S U_{\mu}
\ee
with $R_i$ and $S$ scalar functions and
$V_{i\mu}=\gamma_{\mu},P_{1\mu},P_{2\mu}$ such that 
\be
    (Q\cdot U)R_i Q \cdot V_{i} + S Q \cdot U &=&0  \quad.
\ee
Making use of this condition we construct the remaining vectors 
as a simple extension of the ones proposed by 
K{\i}z{\i}lers\"{u} {\it et. al.}. We list all the vectors below:
\end{itemize}

\be
   &T_{\mu}^{1}&=\left[P_{2\mu}(P_1\cdot Q)-
   P_{1\mu}(P_2\cdot Q)\right] \nonumber\\
   &T_{\mu}^{2}&= T_{\mu}^{1}
   ({\not\! P_1}+{\not\! P_2})\nonumber\\
   &T_{\mu}^{3}&=Q^2\gamma_{\mu}-Q_{\mu}{\not \! Q}\nonumber\\
   &T_{\mu}^{4}&=Q^2\left[\gamma_{\mu}({\not \! P_1}+
   {\not \! P_2})-P_{1\mu}-P_{2\mu}\right]-
   2(P_1-P_2)_{\mu}P_1^{\lambda}P_2^{\nu}\sigma_{\lambda\nu}\nonumber\\
   &T_{\mu}^{5}&=Q^{\nu}\sigma_{\nu\mu}\nonumber\\
   &T_{\mu}^{6}&=-\gamma_{\mu}(P_1^2-P_2^2)+(P_1+P_2)_{\mu}
   {\not \! Q}\nonumber\\
   &T_{\mu}^{7}&=-\frac{1}{2}(P_1^2-P_2^2)
   \left[\gamma_{\mu}({\not \! P_1}+{\not \! P_2})
   -P_{1\mu}-P_{2\mu} \right]
   +(P_1+P_2)_{\mu}P_1^{\lambda}P_2^{\nu}\sigma_{\lambda\nu}\nonumber\\
   &T_{\mu}^{8}&=-\gamma_{\mu}P_1^{\nu}P_2^{\lambda}{\sigma_{\nu\lambda}}
   +P_{1\mu}{\not \! P_2}-P_{2\mu}{\not \! P_1}\nonumber\\
   &T_{\mu}^{i}&= T_{\mu}^{i- {\tiny{8}}} \; {\not \! U} 
   \hspace{1cm}  i=9,16\nonumber \\
   &T_{\mu}^{17}&= (Q \cdot U) \left[P_{2\mu}(P_1\cdot Q)
   +P_{1\mu}(P_2\cdot Q)\right]
   -2 (P_1 \cdot Q) \; (P_2 \cdot Q) \, U_{\mu}\nonumber \\
   &T_{\mu}^{18}&=T_{\mu}^{17} \, ({\not \! P_1}+{\not \! P_2}) \, 
   {\not \! U}\nonumber \\
   &T_{\mu}^{19}&=  \left[ \, (Q \cdot U) \, 
   ( Q^2\gamma_{\mu}+Q_{\mu}{\not \! Q} ) 
   - 2 Q^2 {\not \! Q} \, U_{\mu} \, \right] {\not \! U}\nonumber \\ 
   &T_{\mu}^{20}&= (Q \cdot U) \left[ Q^2 \left\{ 
   \gamma_{\mu}({\not \! P_1}+{\not \! P_2}) + P_{1\mu}+P_{2\mu}
   \right\}+2Q_{\mu}P_1^{\lambda}P_2^{\nu}\sigma_{\lambda\nu}  
   \right] {\not \! U}\nonumber \\
   &&- \; U_{\mu} \left[  Q^2 \left\{ 
   {\not \! Q}({\not \! P_1}+{\not \! P_2})+ P_1^2 - P_2^2 \right\} 
   \; \, + 2 Q^2 P_1^{\lambda}P_2^{\nu}\sigma_{\lambda\nu}  
   \right]{\not \! U}\nonumber \\
   &T_{\mu}^{21}&= \left[ (Q \cdot U) \, Q_{\mu} - Q^2 U_{\mu} 
   \right] {\not \! U}\nonumber \\
   &T_{\mu}^{22}&= (Q \cdot U) \, \left[ \gamma_{\mu} (P_1^2-P_2^2) 
   + (P_1+P_2)_{\mu} {\not \! Q} \right] - 2 (P_1^2-P_2^2) {\not \! Q} 
   U_{\mu}\nonumber \\
   &T_{\mu}^{23}&=  (Q \cdot U) \left[ \frac{1}{2} \, (P_1^2-P_2^2) 
   \left\{ \gamma_{\mu}({\not \! P_1}+{\not \! P_2}) + P_{1\mu}+P_{2\mu}
   \right\}+ (P_1+P_2)_{\mu} P_1^{\lambda}P_2^{\nu}\sigma_{\lambda\nu}  
   \right]\nonumber \\
   &&- \; U_{\mu} \left[  \frac{1}{2} (P_1^2-P_2^2) \left\{ 
   {\not \! Q}({\not \! P_1}+{\not \! P_2})+P_1^2-P_2^2 \right\} 
   \, \, +  (P_1^2-P_2^2) P_1^{\lambda}P_2^{\nu}\sigma_{\lambda\nu}  
   \right]\nonumber \\
   &T_{\mu}^{24}&= (Q \cdot U) \, \left[ P_{2\mu} {\not \! P_1} - 
   P_{1\mu} {\not \! P_2} - \gamma_{\mu} \,  
   P_1^{\lambda}P_2^{\nu}\sigma_{\lambda\nu}  \right]\nonumber\\ 
   && - U_{\mu} \, \left[  (Q \cdot P_2)  {\not \! P_1} - (Q \cdot P_1) 
   {\not \! P_2} -  {\not \! Q}  \, 
   P_1^{\lambda}P_2^{\nu}\sigma_{\lambda\nu} \right]\, ,
   \label{transvbasis}
\ee
where we define $\sigma_{\mu\nu}$ as
\be
   \sigma_{\mu\nu}=\frac{1}{2}[\gamma_{\mu},\gamma_{\nu}]\, .
   \label{sig}
\ee

\section{Covariant HTL vertex and coefficients of the transverse
         basis}\label{IV} 

With the choice of the transverse basis given in
Eq.~(\ref{transvbasis}), the computation of the coefficients $\tau^i$
is a straightforward though lengthy exercise. The first step consists of 
finding the
explicit covariant expression for $\Gamma_\mu$. Then a 
subtraction of the longitudinal vertex $\Gamma_\mu^L$, given by
Eq.~(\ref{long}) yields the transverse vertex. 

Computation of $\Gamma_\mu$ requires finding the different
components of the function $G_{\mu\nu}$, defined in
Eq.~(\ref{funcG}). These have already been found in terms of a
projection onto a particular choice of vector and tensor structures in
Ref.~\cite{Frenkel}. However, in order to explicitly preserve the
symmetric properties of the different elements that make up the vertex
$\Gamma_\mu$ during the computation, we find it convenient to modify
the choice of vector and tensor structures with respect to that of
Ref.~\cite{Frenkel}. 
In this section, we list the results for the components of the
function $G_{\mu\nu}$, leaving a brief sketch of their computation for
the appendix. Let us start with $G_{00}$ which is explicitly given by
\be
   G_{00}(P_1,P_2)\equiv M(P_1,P_2)=\left\{
   \begin{array}{cc}
         \frac{1}{2\sqrt{-\Delta}} \ln\left[\frac{P_1\cdot P_2+\sqrt{-\Delta}}
         {P_1\cdot P_2-\sqrt{-\Delta}}\right] & \Delta <0\\
         \frac{1}{P_1\cdot P_2} & \Delta =0\\
         \frac{1}{\sqrt{\Delta}}\tan^{-1}\left[\frac{\sqrt{\Delta}}
         {P_1\cdot P_2}\right] & \Delta > 0
\end{array}
\right.\, ,
\label{G00}
\ee
where we have defined the quantity $\Delta$ as
\be
   \Delta\equiv P_1^2P_2^2-(P_1\cdot P_2)^2\, .
   \label{delta}
\ee
For the computation of $G_{0i}$, we introduce the set of orthonormal
vectors ${\mathbf \hat{n}}$, ${\mathbf \hat{l}}_+$ and 
${\mathbf \hat{l}}_-$ given in terms of ${\mathbf p}_1$ and 
${\mathbf p}_2$ by
\be
   {\mathbf \hat{n}}&=& \frac{1}{\sqrt{\delta}} 
    \;\left[{\mathbf \hat{p}}_1\times{\mathbf \hat{p}}_2 \right]\nonumber\\
   {\mathbf \hat{l}}^\pm&=&\frac{1}{\sqrt{2\delta_\pm}} \; 
   \left[ {\mathbf \hat{p}}_1\pm  
   {\mathbf \hat{p}}_2  \right] \,,
%   \label{nll}
\ee
where
\be
  \delta_\pm  &\equiv&  1\pm({\mathbf \hat{p}}_1\cdot{\mathbf
   \hat{p}}_2) \nonumber\\
 \delta &\equiv& 1-({\mathbf \hat{p}}_1\cdot{\mathbf \hat{p}}_2)^2
=  \delta_+\delta_-     \nonumber\\
   \delta_{ij}&=&{\mathbf \hat{n}}_i{\mathbf \hat{n}}_j
   +{\mathbf \hat{l}}^+_i{\mathbf \hat{l}}^+_j
   +{\mathbf \hat{l}}^-_i{\mathbf \hat{l}}^-_j\, .
   \label{properties}
\ee
In terms of these vectors, we can express $G_{0i}$ as
\be
   G_{0i}(P_1,P_2)=A{\mathbf \hat{n}}_i+
   B^+{\mathbf \hat{l}}^+_i+B^-{\mathbf \hat{l}}^-_i\, .
   \label{G0i}
\ee
The coefficient $A$ is found by contracting Eq.~(\ref{G0i}) with
${\mathbf \hat{n}}^i$ whereas $B^\pm$ are found upon contraction with
${\mathbf \hat{l}}_\pm^i$, respectively. They are explicitly given by 
\be
   A&=&0\nonumber\\
   B^\pm&=&\frac{1}{\sqrt{2\delta_\pm}}\left[
   \left(\frac{p_{10}}{p_1}\pm\frac{p_{20}}{p_2}\right)M(P_1,P_2)
   -\left(\frac{L(P_2)}{p_1}\pm\frac{L(P_1)}{p_2}\right)\right]\, .
   \label{AB}
\ee
Finally, $G_{ij}$ can be written as
\be
   G_{ij}(P_1,P_2)=X \, {\mathbf \hat{n}}_i{\mathbf \hat{n}}_j \; + \; 
   \tilde{Y}^+ \, {\mathbf \hat{l}}^+_i{\mathbf \hat{l}}^+_j   \; + \;
   \tilde{Y}^- \, {\mathbf \hat{l}}^-_i{\mathbf \hat{l}}^-_j   \; + \;
   Z \, (\, {\mathbf \hat{l}}^+_i{\mathbf \hat{l}}^-_j+
   {\mathbf \hat{l}}^-_i{\mathbf \hat{l}}^+_j \, )\, ,
   \label{Gij}
\ee
where the coefficients $\tilde{Y}^\pm$ and $Z$ are found by contracting
Eq.~(\ref{Gij}) with ${\mathbf \hat{l}}^\pm_i{\mathbf \hat{l}}^\pm_j$
and ${\mathbf \hat{l}}^+_i{\mathbf \hat{l}}^-_j$, respectively,
whereas $X$ can be found  by taking the
trace of  Eq.~(\ref{Gij}). They are explicitly given by
\be
   X&=&\frac{1}{p_1^2p_2^2\delta}\left[
   \Delta M(P_1,P_2)+{\mathbf p}_1\cdot
   (p_{20}{\mathbf p}_1-p_{10}{\mathbf p}_2)L(P_1)+
   {\mathbf p}_2\cdot
   (p_{10}{\mathbf p}_2-p_{20}{\mathbf p}_1)
   L(P_2)\right]\footnotemark\nonumber\\ 
   \tilde{Y}^\pm&=&\frac{1}{2\delta_\pm}
   \left(\frac{p_{10}}{p_1}\pm\frac{p_{20}}{p_2}\right)\left[
   \left(\frac{p_{10}}{p_1}\pm\frac{p_{20}}{p_2}\right)M(P_1,P_2)
   -\left(\frac{L(P_2)}{p_1}\pm\frac{L(P_1)}{p_2}\right)\right]
   \nonumber\\
   && \pm
   \frac{1}{2p_1p_2}\left[2-p_{10}L(P_1)-p_{20}L(P_2)
   \right]\nonumber\\
   Z&=&\frac{1}{2\sqrt{\delta}}\left[
   \left(\frac{p_{10}^2}{p_1^2}-\frac{p_{20}^2}{p_2^2}\right)M(P_1,P_2)-
   \left(p_{10}+\frac{p_{20}}{p_2^2}{\mathbf p}_1\cdot
   {\mathbf p}_2\right)\frac{L(P_2)}{p_1^2}\right.\nonumber\\
   && \hspace{12mm} + \left.
   \left(p_{20}+\frac{p_{10}}{p_1^2}{\mathbf p}_1\cdot
   {\mathbf p}_2\right)\frac{L(P_1)}{p_2^2}\right]\, .
   \label{XYZ}
\ee 
\footnotetext{We find a sign difference for the term $\Delta
              M(P_1,P_2)$ in the expression for the coefficient $X$,
              as compared to the corresponding expression in
              Ref.~\cite{Frenkel}.} 
\noindent
In order to obtain the explicit vectors in terms of which the vertex
$\Gamma_\mu$ is expressed, we first look at the space-like part, namely
$\Gamma_i=m_f^2G_{i\nu}\gamma^\nu$ and add the necessary time-like
pieces to form the corresponding four-vectors. We then subtract the
above added pieces which, together with the time-like part of the
vertex, namely $\Gamma_0=m_f^2G_{0\nu}\gamma^\nu$, become the
terms proportional to the vector $U_\mu$. We finally obtain
\be
   \Gamma_\mu(P_1,P_2)=-m_f^2\left\{X\gamma_\mu+{\mathcal C}_1P_{1\mu}
   +{\mathcal C}_2P_{2\mu}+{\mathcal D}U_\mu\right\}\, ,
   \label{Gammavec}
\ee
where the functions ${\mathcal C}_1$, ${\mathcal C}_2$ and
${\mathcal D}$ are given by
\be
   {\mathcal C}_1&=&-\frac{\gamma_0}{p_1}\left(\frac{B^+}{\sqrt{2\delta_+}}+
   \frac{B^-}{\sqrt{2\delta_-}}\right)+
  \frac{{\mathbf p}_1\cdot\mbox{\boldmath $\gamma$}}{p_1^2}
    \left(\frac{Y^+}{2\delta_+}+
   \frac{Y^-}{2\delta_-}+
   \frac{Z}{\sqrt{\delta}}\right) +
    \frac{{\mathbf p}_2\cdot\mbox{\boldmath $\gamma$}}{p_1p_2}
   \left(\frac{Y^+}{2\delta_+}-
   \frac{Y^-}{2\delta_-}\right)  
   \nonumber\\
   {\mathcal C}_2&=&-\frac{\gamma_0}{p_2}\left(\frac{B^+}{\sqrt{2\delta_+}}-
   \frac{B^-}{\sqrt{2\delta_-}}\right)+
    \frac{{\mathbf p}_2\cdot\mbox{\boldmath $\gamma$}}{p_2^2}
   \left(\frac{Y^+}{2\delta_+}+
   \frac{Y^-}{2\delta_-}-
    \frac{Z}{\sqrt{\delta}}\right)+
     \frac{{\mathbf p}_1\cdot\mbox{\boldmath $\gamma$}}{p_1p_2}
   \left(\frac{Y^+}{2\delta_+}-
   \frac{Y^-}{2\delta_-}\right)
   \nonumber\\
   {\mathcal D}&=&-(M+X)\gamma_0
  +\frac{{\mathbf p}_1\cdot\mbox{\boldmath $\gamma$}}{p_1}
   \left(\frac{B^+}{\sqrt{2\delta_+}}+
   \frac{B^-}{\sqrt{2\delta_-}}\right) +
   \frac{ {\mathbf p}_2\cdot\mbox{\boldmath 
   $\gamma$}}{p_2}\left(\frac{B^+}{\sqrt{2\delta_+}}-
   \frac{B^-}{\sqrt{2\delta_-}}\right)
  -{\mathcal C}_1p_{10}-{\mathcal C}_2p_{20}\, ,
   \label{CD}
\ee
and we have defined $Y^\pm=\tilde{Y}^\pm - X$. In order to obtain the
spin structures out of Eqs.~(\ref{Gammavec}) 
and~(\ref{CD}), we notice that, working in a general frame, we can write
\be
   \gamma_0&=&U\dia\nonumber\\
   {\mathbf p}\cdot\mbox{\boldmath $\gamma$}&=&
   p_0U\dia - P\dia\, .
   \label{fromframe}
\ee  
Making use of these identities and of Eq.~(\ref{CD}), we can write out
Eq.~(\ref{Gammavec}) in terms of the basis vectors given in 
Eq.~(\ref{vbasis}):
\be
   \Gamma_{\mu}(P_1,P_2) &=& -m_f^2 \; \left[  
{\tilde{h}}_{1} V_{1\mu} + {\tilde{h}}_{2} V_{2\mu} + {\tilde{h}}_{3} 
V_{3\mu} + 
{\tilde{h}}_{4} V_{4\mu} + {\tilde{h}}_{6} V_{6\mu} 
+ {\tilde{h}}_{19} V_{19\mu} + 
{\tilde{h}}_{20} V_{20\mu} + {\tilde{h}}_{26} V_{26\mu} 
+ {\tilde{h}}_{27} V_{27\mu} + {\tilde{h}}_{28} V_{28\mu} 
 \right],  \label{HTLvertex}  \nonumber \\
\ee
where
\be
  \tilde{h}_1&=&-\frac{1}{p_1^2}\left(
   \frac{Y^+}{2\delta_+}+
   \frac{Y^-}{2\delta_-}+
   \frac{Z}{\sqrt{\delta}}\right)
   \nonumber\\
   \tilde{h}_2&=&-\frac{1}{p_2^2}\left(
   \frac{Y^+}{2\delta_+}+
   \frac{Y^-}{2\delta_-}-
   \frac{Z}{\sqrt{\delta}}\right)
   \nonumber\\
   \tilde{h}_3&=&\tilde{h}_4=-\frac{1}{p_1p_2}\left(
   \frac{Y^+}{2\delta_+}-
   \frac{Y^-}{2\delta_-}\right)
   \nonumber\\
   \tilde{h}_6&=&X
   \nonumber\\
   \tilde{h}_{19}&=&-\frac{1}{p_1}
   \left(\frac{B^+}{\sqrt{2\delta_+}}+
   \frac{B^-}{\sqrt{2\delta_-}}\right)+
   \frac{p_{10}}{p_1^2}
   \left(\frac{Y^+}{2\delta_+}+
   \frac{Y^-}{2\delta_-}+
   \frac{Z}{\sqrt{\delta}}\right)+
   \frac{p_{20}}{p_1p_2}
   \left(\frac{Y^+}{2\delta_+}-
   \frac{Y^-}{2\delta_-}\right)
   \nonumber\\
   \tilde{h}_{20}&=&-\frac{1}{p_2}
   \left(\frac{B^+}{\sqrt{2\delta_+}}-
   \frac{B^-}{\sqrt{2\delta_-}}\right)+
   \frac{p_{20}}{p_2^2}
   \left(\frac{Y^+}{2\delta_+}+
   \frac{Y^-}{2\delta_-}-
   \frac{Z}{\sqrt{\delta}}\right)+
   \frac{p_{10}}{p_1p_2}
   \left(\frac{Y^+}{2\delta_+}-
   \frac{Y^-}{2\delta_-}\right)
   \nonumber\\
   \tilde{h}_{26}&=&M-3X-\frac{{p_{10}}^2}{p_1^2}\left(
   \frac{Y^+}{2\delta_+}+
   \frac{Y^-}{2\delta_-}+
   \frac{Z}{\sqrt{\delta}}\right)-
\frac{{p_{20}}^2}{p_2^2}\left(
   \frac{Y^+}{2\delta_+}+
   \frac{Y^-}{2\delta_-}-
   \frac{Z}{\sqrt{\delta}}\right) -
   \frac{2 p_{10} p_{20}}{p_1 p_2}
    \left(\frac{Y^+}{2\delta_+}-
   \frac{Y^-}{2\delta_-}\right)
   \nonumber\\
   \tilde{h}_{27}&=&-\frac{1}{p_1}
   \left(\frac{B^+}{\sqrt{2\delta_+}}+
   \frac{B^-}{\sqrt{2\delta_-}}\right)+
   \frac{p_{10}}{p_1^2}
   \left(\frac{Y^+}{2\delta_+}+
   \frac{Y^-}{2\delta_-}+
   \frac{Z}{\sqrt{\delta}}\right)+
   \frac{p_{20}}{p_1p_2}
   \left(\frac{Y^+}{2\delta_+}-
   \frac{Y^-}{2\delta_-}\right)
   \nonumber\\
   \tilde{h}_{28}&=&-\frac{1}{p_2}
   \left(\frac{B^+}{\sqrt{2\delta_+}}-
   \frac{B^-}{\sqrt{2\delta_-}}\right)+
   \frac{p_{20}}{p_2^2}
   \left(\frac{Y^+}{2\delta_+}+
   \frac{Y^-}{2\delta_-}-
   \frac{Z}{\sqrt{\delta}}\right)+
   \frac{p_{10}}{p_1p_2}
   \left(\frac{Y^+}{2\delta_+}-
   \frac{Y^-}{2\delta_-}\right)\, .
   \label{htildes}
\ee
Eqs.~(\ref{HTLvertex}) and~(\ref{htildes}) complete the covariantization
of the full QED vertex at one loop in HTL approximation. The coefficients
${\tilde{h}}_{5}, {\tilde{h}}_{7 \cdots 18}, {\tilde{h}}_{21 \cdots 25}$ and
${\tilde{h}}_{29 \cdots 32}$ of the vectors $V_{5\mu}, V_{7\mu \cdots 18\mu},
V_{21\mu \cdots 25\mu}$ and $V_{29\mu \cdots 32\mu}$ respectively are
zero at this order. Subtracting the longitudinal 
vertex, Eq.~(\ref{longHTL}),  the transverse
vertex $\Gamma_\mu^T$ comes out to be
\be
   \Gamma_\mu^T&=&\Gamma_\mu-\Gamma_\mu^L\nonumber\\
   &=&-m_f^2\Big\{h_1 V_{1\mu} + h_2 V_{2\mu} +
   h_3 V_{3\mu} + h_4 V_{4\mu} + h_6 V_{6\mu}  + h_{19} V_{19\mu} +
   h_{20} V_{20\mu} + h_{26} V_{26\mu} + h_{27} V_{27\mu} 
   +h_{28} V_{28\mu} \Big\}, \nonumber \\
   \label{hs}
\ee
where the coefficients $h_i$  are given by
\be
   h_1&=&  {\tilde{h}}_1 - 
   \frac{\left[\tilde{a}(P_1)-\tilde{a}(P_2)\right]}{2(P_1^2-P_2^2)}
   \nonumber\\
   h_2&=& {\tilde{h}}_2 -
   \frac{\left[\tilde{a}(P_1)-\tilde{a}(P_2)\right]}{2(P_1^2-P_2^2)}
   \nonumber\\
   h_3&=&h_4=   {\tilde{h}}_3 - 
   \frac{\left[\tilde{a}(P_1)-\tilde{a}(P_2)\right]}{2(P_1^2-P_2^2)}
   \nonumber\\
   h_6&=& {\tilde{h}}_6 -\frac{\left[\tilde{a}(P_1)+\tilde{a}(P_2)\right]}{2}
   \nonumber\\
   h_{19}&=&  {\tilde{h}}_{19}
   - \frac{\left[\tilde{b}(P_1)-\tilde{b}(P_2)\right]}{P_1^2-P_2^2}
   \nonumber\\
   h_{20}&=&  {\tilde{h}}_{20}
   -\frac{\left[\tilde{b}(P_1)-\tilde{b}(P_2)\right]}{(P_1^2-P_2^2)}
   \nonumber\\
   h_{26} &=&  {\tilde{h}}_{26}  \nonumber \\
   h_{27}&=&  {\tilde{h}}_{27}  \nonumber \\
   h_{28}&=&   {\tilde{h}}_{28}\, .
   \label{hsexpl}
\ee
By expressing Eq.~(\ref{hs}) in terms of the set of 32 elemental spin
structures and, on the other hand, comparing this with the expansion of
the transverse vertex given by Eq.~(\ref{trsnbas}), we find that the
coefficients $\tau_i$, $i=1\ldots 24$, satisfy the following systems
of linear equations
\be
   (P_2\cdot Q)\tau_2+\tau_3-\tau_6-
   (Q\cdot U)\tau_{22}&=&h_1\nonumber\\
  -(P_1\cdot Q)\tau_2+\tau_3+\tau_6+
   (Q\cdot U)\tau_{22}&=&h_2\nonumber\\
   (P_2\cdot Q)\tau_2-\tau_3+\tau_6-\tau_8+(Q\cdot U)\tau_{22}+
   (Q\cdot U)\tau_{24}&=&h_3\nonumber\\
   -(P_1\cdot Q)\tau_2-\tau_3-\tau_6+\tau_8-(Q\cdot U)\tau_{22}-
   (Q\cdot U)\tau_{24}&=&h_3\nonumber\\
   \tau_8+(Q\cdot U)\tau_{24}&=&0\nonumber\\
   -Q^2\tau_3+(P_1^2-P_2^2)\tau_6-(P_1\cdot P_2)\tau_8-
   (Q\cdot U)(P_1^2-P_2^2)\tau_{22}-(Q\cdot U)(P_1\cdot P_2)
   \tau_{24}&=&h_6\nonumber\\
   2(P_1^2-P_2^2)\tau_{22}+2(Q\cdot P_2)\tau_{24}&=&h_{27}\nonumber\\
   -2(P_1^2-P_2^2)\tau_{22}-2(Q\cdot P_1)\tau_{24}&=&h_{28}\, ,
   \label{set1}
\ee
\be
   -(P_2\cdot Q)\tau_1-(Q^2-2P_1\cdot P_2)\tau_4 + \tau_5 +\frac{1}{2}
   (P_1^2-P_2^2-2P_1\cdot P_2)\tau_7
   &+&\nonumber\\
   (Q\cdot U)(P_2\cdot Q)\tau_{17} +
   \frac{1}{2}(Q\cdot U)(P_1^2-P_2^2-2P_1\cdot P_2)\tau_{23}&=&0
   \nonumber\\
   (P_1\cdot Q)\tau_1-(Q^2+2P_1\cdot P_2)\tau_4 - \tau_5 +\frac{1}{2}
   (P_1^2-P_2^2-2P_1\cdot P_2)\tau_7
   &+&\nonumber\\
   (Q\cdot U)(P_1\cdot Q)\tau_{17} +
   \frac{1}{2}(Q\cdot U)(P_1^2-P_2^2-2P_1\cdot P_2)\tau_{23}&=&0
   \nonumber\\
   -2\tau_4+\tau_7+(Q\cdot U)\tau_{23}&=&0\nonumber\\
   2\tau_4+\tau_7+(Q\cdot U)\tau_{23}&=&0\nonumber\\
   Q^2\tau_4-\tau_5-\frac{1}{2}(P_1^2-P_2^2)\tau_7+\frac{1}{2}
   (Q\cdot U)(P_1^2-P_2^2)\tau_{23}&=&0\nonumber\\
   Q^2\tau_4+\tau_5-\frac{1}{2}(P_1^2-P_2^2)\tau_7+\frac{1}{2}
   (Q\cdot U)(P_1^2-P_2^2)\tau_{23}&=&0\nonumber\\
   -2(P_1\cdot Q)(P_2\cdot Q)\tau_{17}-(P_1^2-P_2^2)
   (P_1^2-P_2^2-2P_1\cdot P_2)\tau_{23}&=&0\nonumber\\
   2(P_1^2-P_2^2)\tau_{23}&=&0\, ,
   \label{set2}
\ee
\be
   -(P_2\cdot Q)\tau_{10}-\tau_{11}+\tau_{14}+(Q\cdot U)
   (P_2\cdot Q)\tau_{18}+(Q\cdot U)\tau_{19}&=&0\nonumber\\
   (P_1\cdot Q)\tau_{10}-\tau_{11}-\tau_{14}+(Q\cdot U)
   (P_1\cdot Q)\tau_{18}+(Q\cdot U)\tau_{19}&=&0\nonumber\\
   -(P_2\cdot Q)\tau_{10}+\tau_{11}-\tau_{14}+\tau_{16}
   +(Q\cdot U)(P_2\cdot Q)\tau_{18}-(Q\cdot U)\tau_{19}&=&0\nonumber\\
   (P_1\cdot Q)\tau_{10}+\tau_{11}+\tau_{14}-\tau_{16}
   +(Q\cdot U)(P_1\cdot Q)\tau_{18}-(Q\cdot U)\tau_{19}&=&0\nonumber\\
   \tau_{16}&=&0\nonumber\\
   Q^2\tau_{11}-(P_1^2-P_2^2)\tau_{14}+(P_1\cdot P_2)\tau_{16}
   +(Q\cdot U)Q^2\tau_{19}&=&0\nonumber\\
   -2(P_1\cdot Q)(P_2\cdot Q)\tau_{18}-2Q^2\tau_{19}&=&0\nonumber\\
   -2(P_1\cdot Q)(P_2\cdot Q)\tau_{18}+2Q^2\tau_{19}&=&0\, ,
   \label{set3}
\ee
\be
   (P_2\cdot Q)\tau_9+(Q^2-2P_1\cdot P_2)\tau_{12}-\tau_{13}
   -\frac{1}{2}(P_1^2-P_2^2-2P_1\cdot P_2)\tau_{15}&-&\nonumber\\
   (Q\cdot U)(Q^2-2P_1\cdot P_2)\tau_{20}-(Q\cdot U)\tau_{21}
   &=&h_{19}\nonumber\\
   -(P_1\cdot Q)\tau_9+(Q^2+2P_1\cdot P_2)\tau_{12}+\tau_{13}
   -\frac{1}{2}(P_1^2-P_2^2-2P_1\cdot P_2)\tau_{15}&-&\nonumber\\
   (Q\cdot U)(Q^2+2P_1\cdot P_2)\tau_{20}+(Q\cdot U)\tau_{21}
   &=&h_{20}\nonumber\\
  -2\tau_{12}+\tau_{15}+2(Q\cdot U)\tau_{20}&=&0\nonumber\\
   2\tau_{12}+\tau_{15}-2(Q\cdot U)\tau_{20}&=&0\nonumber\\
   Q^2\tau_{12}-\tau_{13}-\frac{1}{2}(P_1^2-P_2^2)\tau_{15}+
   (Q\cdot U)Q^2\tau_{20}&=&0\nonumber\\
   Q^2\tau_{12}+\tau_{13}-\frac{1}{2}(P_1^2-P_2^2)\tau_{15}+
   (Q\cdot U)Q^2\tau_{20}&=&0\nonumber\\
   2Q^2(P_1^2-P_2^2-2P_1\cdot P_2)\tau_{20}+Q^2\tau_{21}
   &=&h_{26}\nonumber\\
   4Q^2\tau_{20}&=&0\, .
   \label{set4}
\ee
Equations~(\ref{set1})-(\ref{set4}) constitute 4 sets, each consisting
of 8 linear algebraic equations for the 24 unknowns $\tau_i$. This
means that not all of the equations are independent and therefore,
there must be some relations among the several inhomogeneous terms. It
is also a straightforward exercise to verify that indeed this
is the case and that such relations are
\be
   (P_1\cdot Q)h_{19}+(P_2\cdot Q)h_{20}+(Q\cdot U)h_{26} &=&0\nonumber\\
  (P_1\cdot Q)h_1- (P_2\cdot Q)h_2  - Q^2h_3 + 2h_6 +
   (Q\cdot U)(h_{27}-h_{28})  &=& 0\nonumber\\
  (P_1\cdot Q)h_1+(P_2\cdot Q)h_2+ (P_1^2-P_2^2)h_3+
   (Q\cdot U)(h_{27}+h_{28}) &=&0\, .
\ee
The non-vanishing coefficients which solve the sets of 
Eqs.~(\ref{set1})-(\ref{set4}) are given by
\be
   \tau_2&=&-\frac{1}{2Q^2}(h_1+h_2+2h_3)\nonumber\\
   \tau_3&=& \frac{1}{4} (h_1+h_2-2h_3)    \nonumber\\
   \tau_6&=&-\frac{1}{4}(h_1-h_2)+\frac{(Q\cdot U)}{2Q^2(P_1^2-P_2^2)}
   \left[(P_2\cdot Q)h_{27}+(P_1\cdot Q)h_{28}\right]\nonumber\\
   \tau_8&=&\frac{(Q\cdot U)}{Q^2}\left(h_{27}+h_{28}\right)\nonumber\\
   \tau_9&=&-\frac{1}{Q^2}(h_{19}+h_{29})\nonumber\\
   \tau_{21}&=&\frac{1}{Q^2}h_{26}\nonumber\\
   \tau_{22}&=&\frac{1}{2Q^2(P_1^2-P_2^2)}
   \left[(P_1\cdot Q)h_{27}+(P_2\cdot Q)h_{28}\right]\nonumber\\
   \tau_{24}&=&-\frac{1}{2Q^2}(h_{27}+h_{28})\, .
   \label{tausexp}
\ee
We would like to emphasize the following points:

\begin{itemize}

\item The ${\tau_i}$ in Eqs.~(\ref{tausexp}) are
independent of the gauge parameter ${\xi}$, since they are obtained
from the HTL vertex, which is known to be gauge fixing independent.  

\item  Given that, at finite temperature in the HTL approximation, the 
explicit expression for the vertex $\Gamma_{\mu}(P_1,P_2)$, 
Eqs.~(\ref{Vert}) and~(\ref{funcG}), is symmetric under the exchange of
$P_1$ and $P_2$, the coefficients of (anti)symmetric vectors $T^{i}_{\mu}$
are also (anti)symmetric.

\end{itemize}
Note that $\tau_6$ and $\tau_{22}$ have a kinematic singularity
when $P_1^2 \rightarrow P_2^2$. In this limit, these vectors point along 
the same direction
\be
  T_{\mu}^{6}&=&-\gamma_{\mu}(P_1^2-P_2^2)+(P_1+P_2)_{\mu}
   {\not \! Q} \nonumber\\
  &\stackrel{\scriptscriptstyle P_1^2\rightarrow P_2^2}{\longrightarrow}&
  {\not \! Q}(P_1+P_2)_{\mu} \nonumber\\
  T_{\mu}^{22}&=&(Q \cdot U) \, \left[ \gamma_{\mu} (P_1^2-P_2^2) 
   + (P_1+P_2)_{\mu} {\not \! Q} \right] - 2 (P_1^2-P_2^2) {\not \! Q} 
   U_{\mu}\nonumber\\
  &\stackrel{\scriptscriptstyle P_1^2\rightarrow P_2^2}{\longrightarrow}&
  {\not \! Q}(Q \cdot U) \, (P_1+P_2)_{\mu}\, .
  \label{parallel} 
\ee
Therefore, they can 
be added up and the kinematic singularities are canceled out 
in the sum, as expected. This observation helps us redefine 
one of these vectors in such a way that the coefficients of both of them
are independently free of the singularities. 
\be
T_{\mu}^{22} \longrightarrow T_{\mu}^{\prime \, 22} &=& 
        \frac{1}{2(P_1^2-P_2^2)} \, \left[ T_{\mu}^{22} -
         (Q \cdot U) \, T_{\mu}^{6} \right]  \nn \\
     &=&  (Q \cdot U) \gamma_{\mu}- Q\dia U_\mu\, .
    \label{newT22} 
\ee
With this redefinition, the new coefficients $\tau_6^{\prime}$ and 
$\tau_{22}^{\prime}$ are
\be
   \tau_6^{\prime} &=& \tau_6 + (Q \cdot U) \, \tau_{22}  \nn  \\
    &=& -\frac{1}{4} \, (h_1-h_2) \, + \, \frac{(Q \cdot U)}{2 Q^2} \,
        \left[ h_{27} + h_{28} \right]  \nn \\
 \tau_{22}^{\prime} &=& 2(P_1^2-P_2^2) \tau_{22} \nn \\
   &=& \frac{1}{Q^2} \, \left[ (P_1 \cdot Q) h_{27} + (P_2 \cdot Q)
     h_{28}  \right]\, ,   \label{newtaus}
\ee
which are explicitly free of kinematic singularities.

\section{Conclusions}\label{V}

In this paper, we present the one loop calculation of the
fermion-boson vertex in QED at finite temperature in the HTL approximation.  
In the most general form, the vertex can be written in terms of 
32 independent Lorentz vectors. Following the procedure outlined by 
Ball and Chiu, 8 of the 32 vectors define the longitudinal vertex
which satisfies the WTI  relating it to the 
fermion propagator. The transverse vertex is written in terms of the
remaining 24 vectors. The choice of these basis vectors is not unique
but is a natural and straightforward extension of the $T=0$ basis
given in Ref.~\cite{KRP}. We have evaluated the coefficients of the basis
vectors. Given that in this scheme, a kinematic singularity appears in
two of the coefficients, we have modified the choice of one of the
transverse vectors to simultaneously remove both singularities. Any
non-perturbative vertex {\em ansatz} should reproduce
Eqs.~(\ref{tausexp}) and~(\ref{newtaus}) in the weak coupling and high
temperature regime. Therefore, Eqs.~(\ref{tausexp}) 
and~(\ref{newtaus}) should serve as a guide to construct a 
non-perturbative vertex in QED at finite, albeit large, temperature. 
Finally, it is important to also note that the present analysis is
valid for any SU(N) gauge theory, (for instance QCD) provided
that the thermal fermion mass includes the corresponding group
factor~\cite{LeBellac}.

A reliable  non-perturbative vertex is essential for the 
study of dynamical symmetry breaking through the corresponding 
Schwinger-Dyson Equations. One may also be able to make useful 
predictions for the gauge theory contributions to the top-quark
condensate scheme of electroweak symmetry breaking at finite
temperature~\cite{BRZ}, leading to a better insight into the
electroweak phase transition. All this is for the future.

\section*{Acknowledgements}

The authors are in debt to A. Weber for useful discussions and comments.
Support for this work has been received in part by the CIC and
CONACyT under grants numbers 4.12, 32395-E and 32279-E, respectively.

\section*{Appendix}

Here, we explicitly compute the integrals involved in the calculations
of Sect.~\ref{IV}. Capital letters are used throughout to denote
four-momentum vectors. $\hat{K}^\mu=(-1,{\mathbf\hat{k}})$. We start with the
function $G_{00}$
\be
   G_{00}(P_1,P_2)=\int\frac{d\Omega}{4\pi}\frac{1}{(P_1\cdot\hat{K})
   (P_2\cdot\hat{K})}\, .
   \label{A1}
\ee
Introducing the Feynman parameter $x$, we can write
\be
   G_{00}(P_1,P_2)&=&\int_0^1dx\int\frac{d\Omega}{4\pi}
   \frac{1}{\left[(xP_1+(1-x)P_2)\cdot\hat{K}\right]^2}\nonumber\\
   &=&\frac{1}{2}\int_0^1dx\int_{-1}^1d(\cos\theta)
   \frac{1}{[u+v\cos\theta]^2}\, ,
   \label{A2}
\ee
where we have defined the functions $u$ and $v$ by
\be
   u&\equiv&xp_{10}+(1-x)p_{20}\nonumber\\
   v&\equiv&|x{\mathbf p}_1+(1-x){\mathbf p}_2|\, .
   \label{A3}
\ee
After integration over the polar angle $\theta$, we obtain
\be
   G_{00}(P_1,P_2)&=&\int_0^1\frac{dx}{u^2-v^2}\nonumber\\
   &=&\int_0^1\frac{dx}{[xP_1+(1-x)P_2]^2}\, ,
   \label{A4}
\ee
from where the result in Eq.~(\ref{G00}) follows.
Next, for the computation of the function $G_{0i}$, we require to
compute integrals of the type
\be
   \int\frac{d\Omega}{4\pi}\frac{{\mathbf p}_1\cdot{\mathbf\hat{k}}}
   {(P_1\cdot\hat{K})(P_2\cdot\hat{K})}\, .
   \label{A5}
\ee
Adding and subtracting $p_{10}$ in the numerator of the integral in
Eq.~(\ref{A5}), we get
\be
   \int\frac{d\Omega}{4\pi}\frac{{\mathbf p}_1\cdot{\mathbf\hat{k}}}
   {(P_1\cdot\hat{K})(P_2\cdot\hat{K})}&=&
   -\int\frac{d\Omega}{4\pi}\left\{
   \frac{p_{10}-{\mathbf p}_1\cdot{\mathbf\hat{k}}-p_{10}}
   {(P_1\cdot\hat{K})(P_2\cdot\hat{K})}\right\}\nonumber\\
   &=&-\int\frac{d\Omega}{4\pi}\frac{1}{(P_2\cdot\hat{K})}
   -\int\frac{d\Omega}{4\pi}\frac{p_{10}}
   {(P_1\cdot\hat{K})(P_2\cdot\hat{K})}\, .
   \label{A6}
\ee
The first of the expressions in the second line of Eq.~(\ref{A6}) is
explicitly given by
\be
   \int\frac{d\Omega}{4\pi}\frac{1}{(P_2\cdot\hat{K})}&=&
   -\frac{1}{2}\int_{-1}^1d(\cos\theta)\frac{1}
   {p_{20}+p_2\cos\theta}\nonumber\\
   &=&-\frac{1}{2p_2}\ln\left(\frac{p_{20}+p_2}{p_{20}-p_2}\right)\nonumber\\
   &\equiv&-L(P_2)\, ,
   \label{A7}
\ee
whereas the second of the terms in the second line of
Eq.~(\ref{A6}) is simply $-p_{10}$ times $G_{00}$. Therefore,
\be
   \int\frac{d\Omega}{4\pi}\frac{{\mathbf p}_1\cdot{\mathbf\hat{k}}}
   {(P_1\cdot\hat{K})(P_2\cdot\hat{K})}=-p_{10}M(P_1,P_2)+L(P_2)\, .
   \label{A8}
\ee
Analogously,
\be
   \int\frac{d\Omega}{4\pi}\frac{{\mathbf p}_2\cdot{\mathbf\hat{k}}}
   {(P_1\cdot\hat{K})(P_2\cdot\hat{K})}=-p_{20}M(P_1,P_2)+L(P_1)\, ,
   \label{A9}
\ee
from where the result in Eqs.~(\ref{AB}) follows. Finally, for the
computation of the function $G_{ij}$, we first compute an integral of the type
\be
   \int\frac{d\Omega}{4\pi}\frac{({\mathbf p}_1\cdot{\mathbf\hat{k}})^2}
   {(P_1\cdot\hat{K})(P_2\cdot\hat{K})}\, .
   \label{A10}
\ee
Adding and subtracting $p_{10}$ in one of the factors in the numerator
of Eq.~(\ref{A10}), we obtain
\be
   \int\frac{d\Omega}{4\pi}\frac{({\mathbf p}_1\cdot{\mathbf\hat{k}})^2}
   {(P_1\cdot\hat{K})(P_2\cdot\hat{K})}&=&
   \int\frac{d\Omega}{4\pi}\left\{
   \frac{(p_{10}+{\mathbf p}_1\cdot{\mathbf\hat{k}}-p_{10})
   ({\mathbf p}_1\cdot{\mathbf\hat{k}})}
   {(P_1\cdot\hat{K})(P_2\cdot\hat{K})}\right\}\nonumber\\
   &=&-\int\frac{d\Omega}{4\pi}
   \frac{{\mathbf p}_1\cdot{\mathbf\hat{k}}}{(P_2\cdot\hat{K})}
   +p_{10}\int\frac{d\Omega}{4\pi}
   \frac{1}{(P_2\cdot\hat{K})}\nonumber\\
   &+&p_{10}^2\int\frac{d\Omega}{4\pi}
   \frac{1}{(P_1\cdot\hat{K})(P_2\cdot\hat{K})}\, .
   \label{A11}
\ee
The first of the integrals in the second line of Eq.~(\ref{A11}) is
given explicitly by
\be
   \int\frac{d\Omega}{4\pi}
   \frac{{\mathbf p}_1\cdot{\mathbf\hat{k}}}{(P_2\cdot\hat{K})}&=&
   -\frac{{\mathbf p}_1\cdot{\mathbf p}_2}{2p_2}
   \int_{-1}^1d(\cos\theta)\frac{\cos\theta}{p_{20}+p_2\cos\theta}
   \nonumber\\
   &=&-\frac{{\mathbf p}_1\cdot{\mathbf p}_2}{p_2^2}
   +\left(\frac{p_{20}}{p_2^2}{\mathbf p}_1\cdot
   {\mathbf p}_2\right)L(P_2)\, .
   \label{A12}
\ee
On the other hand, the second of the terms in the second line of
Eq.~(\ref{A11}) is simply $-p_{10}$ times $L(P_2)$, whereas the
third term is just $p_{10}^2$ times $G_{00}$. Therefore, the integral
under scrutiny is given by
\be
   \int\frac{d\Omega}{4\pi}\frac{({\mathbf p}_1\cdot{\mathbf\hat{k}})^2}
   {(P_1\cdot\hat{K})(P_2\cdot\hat{K})}=
   \frac{{\mathbf p}_1\cdot{\mathbf p}_2}{p_2^2}
   -\left(\frac{p_{20}}{p_2^2}{\mathbf p}_1\cdot
   {\mathbf p}_2\right)L(P_2)-p_{10}L(P_2)+p_{10}^2M(P_1,P_2)\, .
   \label{A13}
\ee
Analogously,
\be
   \int\frac{d\Omega}{4\pi}\frac{({\mathbf p}_2\cdot{\mathbf\hat{k}})^2}
   {(P_1\cdot\hat{K})(P_2\cdot\hat{K})}=
   \frac{{\mathbf p}_1\cdot{\mathbf p}_2}{p_1^2}
   -\left(\frac{p_{10}}{p_1^2}{\mathbf p}_1\cdot
   {\mathbf p}_2\right)L(P_1)-p_{20}L(P_1)+p_{20}^2M(P_1,P_2)\, .
   \label{A14}
\ee
The last integral required is given by
\be
   \int\frac{d\Omega}{4\pi}
   \frac{({\mathbf p}_1\cdot{\mathbf\hat{k}})
   ({\mathbf p}_2\cdot{\mathbf\hat{k}})}
   {(P_1\cdot\hat{K})(P_2\cdot\hat{K})}\, ,
   \label{A15}
\ee
Adding and subtracting  $p_{10}$ in the first and $p_{20}$ in the second of 
the factors in the numerator of the above integral, we get
\be
   \int\frac{d\Omega}{4\pi}
   \frac{({\mathbf p}_1\cdot{\mathbf\hat{k}})
   ({\mathbf p}_2\cdot{\mathbf\hat{k}})}
   {(P_1\cdot\hat{K})(P_2\cdot\hat{K})}&=&
   \int\frac{d\Omega}{4\pi}
   \frac{(p_{10}+{\mathbf p}_1\cdot{\mathbf\hat{k}}-p_{10})
   (p_{20}+{\mathbf p}_2\cdot{\mathbf\hat{k}}-p_{20})}
   {(P_1\cdot\hat{K})(P_2\cdot\hat{K})}\nonumber\\
   &=&1+p_{10}\int\frac{d\Omega}{4\pi}
   \frac{1}{(P_1\cdot\hat{K})}+p_{20}\int\frac{d\Omega}{4\pi}
   \frac{1}{(P_2\cdot\hat{K})}\nonumber\\
   &+&p_{10}p_{20}\int\frac{d\Omega}{4\pi}\frac{1}{(P_1\cdot\hat{K})
   (P_2\cdot\hat{K})}\, ,
   \label{A16}
\ee
which, by using Eqs.~(\ref{A1}) and~(\ref{A7}), can be written as
\be
   \int\frac{d\Omega}{4\pi}
   \frac{({\mathbf p}_1\cdot{\mathbf\hat{k}})
   ({\mathbf p}_2\cdot{\mathbf\hat{k}})}
   {(P_1\cdot\hat{K})(P_2\cdot\hat{K})}=
   1-p_{10}L(P_1)-p_{20}L(P_2)+p_{10}p_{20}M(P_1,P_2)\, ,
   \label{A17}
\ee
from where Eqs.~(\ref{XYZ}) follow.

\end{document}